\documentclass[letterpaper,english,reprint, aps]{revtex4-1}
\usepackage[T1]{fontenc}
\usepackage[latin9]{inputenc}
\usepackage{amsmath}
\usepackage{amssymb}
\usepackage{esint}

\makeatletter

\pdfpageheight\paperheight
\pdfpagewidth\paperwidth

\makeatother

\usepackage{babel}
\begin{document}

\preprint{APS/123-QED}

\title{Structural Derivative Model for Tissue Radiation Response}

\author{José Weberszpil}

\email{josewebe@ufrrj.br, josewebe@gmail.com}

\affiliation{Programa de Pós Graduação em Modelagem Matemática e Computacional-PPGMMC,
Universidade Federal Rural do Rio de Janeiro, UFRRJ-IM/DTL, Av. Governador
Roberto Silveira s/n, Nova Iguaçu, Rio de Janeiro, RJ, Brazil.}

\author{Oscar Sotolongo-Costa}

\email{osotolongo@uaem.mx}

\affiliation{Centro de Investigación en Ciencias, (IICBA), Universidad Autónoma
del Estado de Morelos, Av. Universidad 1001, 62209 Cuernavaca, Morelos,
México.}

\date{\today}
\begin{abstract}
By means of a recently-proposed metric or structural derivative, called
$scale-q-$derivative approach, we formulate differential equation
that models the cell death by a radiation exposure in tumor treatments.
The considered independent variable here is the absorbed radiation
dose $D$ instead of usual time. The survival factor, $F_{S}$, for
radiation damaged cell obtained here is in agreement with the literature
on the maximum entropy principle, as it was recently shown and also
exhibits an excellent agreement with the experimental data. Moreover,
the well-known linear and quadratic models are obtained. With this
approach, we give a step forward and suggest other expressions for
survival factors that are dependent on the complex tumor structure.
\end{abstract}
\maketitle

\section{Introduction}

Recently, one of the authors (J.W.) demonstrated the existence of
a possible relationship between $q-$deformed algebras in two different
contexts of Non-extensive Statistical Mechanics (NESM), namely, the
Tsallis' framework and the Kaniadakis' scenario, with local form of
fractional-derivative operators denned in fractal media, the so-called
Hausdorff derivatives, mapped into a continuous medium with a fractal
measure \citep{Weber2015}.

In addition, to describe complex systems, the $q-$calculus, in a
NESM context, has its formal development based on the definition of
deformed expressions for the logarithm and exponential \citep{Borges2004},
namely, the $q-$logarithm and the $q-$exponential. In such a context,
an interesting algebra emerges and the formalism of a deformed derivative
opens new possibilities for the treatment of complex systems outside
of the thermodynamical spectrum, especially those with fractal or
multi-fractal metrics and presenting long-range dynamical interactions.
The deformation parameter or entropic index, $q$, has an important
place in the description of these complex systems, and describes deviations
from standard Lie symmetries. It also provides the formalism to accommodate
scale invariance to the thermodynamic formalism, for systems with
multi fractal properties. For $q\rightarrow1$, the formalism reverts
to the standard one \citep{Borges2004}. In this context, one of the
authors (O.S.) has shown, in a recent article, that radiobiological
models can be derived, considering the Tsallis' NESM, from the maximum
entropy principle and using a cutoff condition motivated by experimental
clinical results, with excellent agreement \citep{Sotolongo-PRL,Brouers-Sotolongo}.

It is our belief that new conceptions and approaches, such as the
fractional derivatives (non local operators) and the structural derivatives
(local operators; also called deformed or metric or fractal derivatives)
\citep{Structurl Deriv,Structural derivative based on inverse Mittag-Leffler},
may allow us to understand new systems, while also helping us to extend
well-known results. In particular, regarding to the problem studied,
the use of structural derivatives, similarly to the fractional calculus
(FC), allows us to describe and emulate complex dynamics involving
multiple species and environmental variation, without the addition
of explicit terms relating to this complexity in the dynamical equations
describing the system.

Here, we set up a formalism that may yield an effective theory to
model cell radiation absorption, without the use of statistical averages
and without formally imposing any specific nonstandard statistics.
For this purpose, we apply the mathematical tool of structural derivatives
\citep{Structurl Deriv} that includes $q-$derivatives-like, called
$scale-q-$derivative \citep{Nosso variacional} and may also be thought
with other kind of structural derivatives like Hausdorff derivatives
\citep{Balankin2012,Balankin2,Chen2006} and conformable derivatives
\citep{Conformable}.

Hence, structural derivatives and/or FC may be the tools to describe,
in a softer manner, connections between a coarse-grained medium and
dissipation at a certain energy scale.

We claim that the use of deformed or structural derivatives may enable
us to consider the effects of internal times of systems or different
internal absorption rates, such as an ecological time proposed in
Ref. \citep{Ecological Time- Fractional}. This makes sense in the
context of complex systems similar to the open species-environmental
medium interaction. The approach considered here, seems to be applicable
to power-law phenomena, mathematical modeling tumor, biological and
logistic growth \citep{Tumor Growth 1,Logistic Models 0,Growth 2,Tumor Growth 3,Tumor Growth 4,Biological Growth,Tumor Growth 6,Seiichi Sakanoue,Growth},
and other dynamical systems.

\section{Comments on the use of structural derivatives}

We also suggested that the understanding of the results for the anomalous
radiation absorption may be thought in the realm of complexity, mean
life, coarse-graininess, and pseudoparticle concepts. Tumor tissue
is in fact, complex coarse-graininess structures with internal structure
and interacting with media and the radiation.

In a coarse-grained space, a point is not infinitely thin, and here,
this feature is modeled by means of a space in which the generic differential
is not $dx$, but rather $(dx)^{\alpha}$.

The great majority of actual classical systems is nonconservative
but, in spite of that, the most advanced formalisms of classical mechanics
deal only with conservative systems \citep{RW}. Dissipation \citep{Sympletic},
for example, is present even at the microscopic level. There is dissipation
in every non-equilibrium or fluctuating process, including dissipative
tunneling \citep{Cal} and electromagnetic cavity radiation \citep{Sen},
for instance. In \citep{Sympletic}, we adopt that a way to suitably
treat nonconservative systems is through fractional Calculus (FC),
since it can be shown that, for example, a friction force has its
form stemming from a Lagrangian that contains a term proportional
to the fractional derivative, which may be a derivative of any non-integer
order \citep{RW,Alireza,Matheus Lazo,g-factor}.

Since we are dealing with open systems, as we already commented, particles
or, in the case here, tumor cells constituting the tissue to be irradiated,
should in fact be seen as dressed entities or pseudoparticles that
exchange energy with other particles and the environment. The system
composed by particles and their surroundings may be considered nonconservative
due to the possible energy exchange. This energy exchange may be responsible
for the resulting noninteger dimension of space-time, giving rise
then to an effective coarse-grained medium. Notice that this kind
of approach is suitable to treat systems with dissipative forces or
non-holonomic systems since it include the scale in time letting to
consider the effects of internal times of the systems. Since the tissue
radiation absorption is a nonconservative process, all of the mentioned
remark are valid here.

Also, we can make an attempt to connect anomalous diffusion with the
anomalous radiation tissue absorption. This connection also re-enforces
the use of alternative mathematical approaches. As the diseased tissue
is exposed by radiations doses, it changes the way it process the
biochemical interactions, including nutrient and gases diffusion or
abortion, cell membrane permeability, excretion of metabolites, and
so on. The disorder caused by radiation induce a level of disorder
in the dynamical processes, with consequent anomalous diffusion like
behavior. Since the disorder here is local, the approach with a deformed
local operator seem to be easier than the approach of FC used in the
study of anomalous diffusion.

It is our belief that new conceptions and approaches, such as metric
derivatives and fractional derivatives, may allow us to understand
new systems, while also helping us to extend well-known results. In
particular, with regards to the studied problem, the use of deformed
derivatives, similarly to the FC, allows us to describe and emulate
complex dynamics involving multiple tissues and environmental variation,
without the addition of explicit terms relating to this complexity
in the dynamical equations describing the system. In some way, the
formalism proposed here may yield an effective theory for interacting
cell-radiation or anomalous growth/death, without the use of statistical
averages. Hence, metric derivatives may be the tools to describe,
in a softer manner, connections between a coarse-grained medium and
dissipation at a certain energy scale.

Some additional support to this view can be found in a recent article
that considers a thermodynamics formulation with fractal structure
\citep{66}. There, the author considered a Hausdorff dimension and
determined the Lipshitz-Hölder exponent in terms of the entropic index
$q$, connecting the thermofractal concept - considered a class of
thermodynamical systems that present a fractal structure in its thermodynamical
description - with hadrons. Interesting to observe that the referred
author also considered the internal energy of subsystems in such a
way that they behave as particles with an internal structure.

\section{methods}

In this section, we present some mathematical aspects concerning three
different kinds of structural derivative, the Hausdorff derivative,
for connections and clarifying reasons, the $q-$derivative in the
context of NESM and the recently proposed $scale-q-$derivative \citep{Nosso variacional}.

\subsubsection*{Hausdorff Derivative}

A model that maps hydrodynamics continuum flow in a fractal coarse-grained
(fractal porous) space, which is essentially discontinuous in the
embedding Euclidean space, into a continuous flow governed by conventional
partial differential equations was suggested in Ref. \citep{Tarasov2005}.
In a latter work, Balankin and Espinoza \citep{Balankin2012} suggested
that the discontinuous fractal flow in a fractally permeable medium
can be mapped into a fractal continuous flow, which is describable
within a continuum framework, indicating also that the geometric framework
of fractal continuum model is the three-dimensional Euclidean space
with a fractal metric. For more details, we refer the interested reader
to \citep{Balankin2012,Balankin2}. Employing the structural differential
operators \citep{Structurl Deriv} in connection with the Hausdorff
derivative \citep{Chen2006}, one can describe the latter as \citep{Balankin2}:

\begin{eqnarray}
\frac{d^{H}}{dx^{\zeta}}f(x) & = & \lim_{x\rightarrow x'}\frac{f(x^{'})-f(x)}{(x^{'})^{\zeta}-x^{\zeta}}=\nonumber \\
 & = & \left(\frac{x}{l_{0}}+1\right)^{1-\zeta}\frac{d}{dx}f=\nonumber \\
 & = & \frac{l_{0}^{\zeta-1}}{c_{1}}\frac{d}{dx}f=\frac{d}{d^{\zeta}x}f,\label{eq:Haudorff derivative}
\end{eqnarray}
where $l_{0}$ is the lower cutoff along the Cartesian $x-$axis and
the scaling exponent, $\zeta$, characterizes the density of states
along the normal direction to the intersection of the fractal continuum
with the plane, as defined in Ref. \citep{Balankin2}.

\subsubsection*{$q-$derivative in the Non-extensive Context}

Over the recent decades, diverse formalisms have emerged that are
adopted to approach complex systems. Among those, we may quote the
$q-$calculus in the Tsallis' version of NESM, with its undeniable
success whenever applied to a wide class of different systems; Kaniadakis'
approach, based on the compatibility between relativity and thermodynamics;
fractional calculus (FC), that deals with the dynamics of anomalous
transport and other natural phenomena, and also some local versions
of FC that claim to be able to study fractal and multi-fractal spaces
and to describe dynamics in these spaces by means of fractional differential
equations.\\

The non-extensive behavior for certain physical systems are in close
relation with the presence of spatial-temporal long-range interactions
\citep{Rajagopal1996}. It is now well known that, for systems as
those with gravitational interaction or those where the range of interactions
is comparable to system size, with the presence of memory effects,
with fractal, multifractal structures, and so on, the standard statistical
mechanics and thermodynamics seems not to be the adequate one to describe
the dynamics. So, Tsallis \citep{Tsallis 88} has proposed a generalization
called nonextensive statistical mechanics and thermodynamics to explain
many of the so called anomalous phenomena in the context of complexity.
We claim that this may include biological and similar systems, where
complexity is ubiquitous \citep{Sotolongo-PRL,Murray 1993- Predator-Pray}.

In close relation with NESM statistics, the $q-$derivative sets up
a deformed algebra and takes into account that the $q-$exponential
is an eigenfunction of $D_{(q)}$ \citep{Borges2004}. Borges proposed
the following operator for $q$-derivative:

\begin{equation}
{\displaystyle D_{(q)}f(x)\equiv{\displaystyle \lim_{y\to x}\frac{f(x)-f(y)}{x\ominus_{q}y}}}={\displaystyle [1+(1-q)x]\frac{df(x)}{dx}.}\label{eq:q-derivative}
\end{equation}

Here, $\ominus_{q}$ is the deformed difference operator, $x\ominus_{q}y\equiv\frac{x-y}{1+(1-q)y}\qquad(y\ne1/(q-1)).$

The q- Integral has the similar structure of Riemann Improper Integral
\citep{Borges2004}:

\begin{equation}
\int_{a}^{t}f(x)d_{q}x=\intop_{a}^{t}\frac{f(x)}{1+(1-q)x}dx;
\end{equation}
$d_{q}x=\underset{y\rightarrow x}{lim}x\ominus_{q}y=\frac{1}{1+(1-q)x}dx$.

There is also a dual derivative operator $D^{(q)}$, associated with
$D_{(q)}$, defined by \citep{Borges2004}

\begin{equation}
D^{(q)}f(x)\equiv{\displaystyle \lim_{y\to x}\frac{f(x)\ominus_{q}f(y)}{x-y}={\displaystyle \frac{1}{1+(1-q)f(x)}\,\frac{df(x)}{dx},}}\label{eq:Dual deriv Borges}
\end{equation}
and its corresponding dual $q$-integral

\begin{equation}
\int\nolimits ^{(q)}f(x)dx\equiv\int[1+(1-q)f(x)]f(x)\;dx.
\end{equation}

The connection of q-derivative with the Hausdorff derivative was recently
demonstrated\citep{Weber2015}. It results that the relevant parameters
may be connected as: 
\begin{equation}
1-q=\frac{(1-\zeta)}{l_{0}}.
\end{equation}

Consequently, one can see that the deformed $q-$derivative is the
first order expansion of the Hausdorff derivative and that there is
a strong connection between these formalisms by means of a fractal
metric.

Recently, one of the authors (J.W.) has introduced what was named
as $scale-q-$derivative \citep{Nosso variacional}:

\begin{equation}
D_{(q,\lambda)}f(\lambda x)=[1+(1-q)\lambda x]\frac{dy(\lambda x)}{dx},
\end{equation}
in such a way that the q-exponential with a scale $\lambda,$ $e_{q}(\lambda x),$
is the eigenfunction of the differential equation

\begin{equation}
[1+(1-q)\lambda x]\frac{de_{q}(\lambda x)}{dx}=\lambda e_{q}(\lambda x),
\end{equation}
with eigenvalue $\lambda,$ as the reader can verify. Here, $e_{q}(x)=[1+(1-q)x]$
is the $q-$exponential function that appears in the context of non
extensive statistical mechanics \citep{Borges2004}.

\section{The model for survival factor}

One of the advantages for the mapping to a fractal continuum is that
the resulting model in this new space may take the random variable
$x$ as a continuous variable, instead of a discrete one. The consideration
of $D$ as a continuous variable was done also in Ref. \citep{Sotolongo-PRL}.

Let $D_{0}$ be the minimal annihilation dose \citep{Sotolongo-PRL},
where no cell survive. Let us name the dimensionless dose as $x=\frac{D}{D_{0}}$.

To gain some insight, let us consider the simple model, considering
the usual derivative.

Here, $N(x)$ is the number of cells at a given state of the radiation
treatment when a dimensionless dose $x$ is achieved. Let us denote
$\alpha$ as the dimensionless cell death probability after the achievement
of dose $x$. The model is analogous to the nuclear decay problem,
that uses time $t$ as the independent variable. Then, we can write:

\begin{equation}
\frac{dN(x)}{dx}=-\alpha N(x)\label{eq:Nuclear Decay}
\end{equation}

Note that $\alpha dx=p(x)$ is a constant probability density to kill
a unitary cell by a radiation dose.

The well known solution of eq.(\ref{eq:Nuclear Decay}) is $N(x)=N_{0}e^{-\alpha x}.$
So, the survival factor, $\frac{N}{N_{0}}=F_{S},$ can be written
as 
\begin{equation}
F_{S}=e^{-\alpha x},
\end{equation}
that reduces to the well known linear model in Ref. \citep{Tubiana}
if $\alpha=1.$

It is important to note that, since the survival factors $F_{S}$
are statistically independent, the doses $x$ are summable in the
linear model \citep{Tubiana}, that means $F_{S}(x_{1},x_{2})=F_{S}(x_{1}+x_{2})$.
But in the context of the approach of complex systems, considering
that tumors have complex structures, this implies that despite the
$F_{S}$ are yet statistically independent, the doses must be here
$q-$summable, due to non-related spaces. That is, $F_{S}(x_{1},x_{2})=F_{S}(x_{1}\oplus_{q}x_{2}),$
where $x_{1}\oplus_{q}x_{2}=x_{1}+x_{2}+(1-q)x_{1}x_{2}.$ This justifies
the approach of the mapping to a continuous fractal and consequently,
the introduction of a structural derivative like $scale-q$ derivative.

Suppose now that we may proceed to a mapping to a fractal continuum
due to a complex structure of the tumor cells and the derivative had
to be redefined as a scale-q-derivative. The differential equation
is now 
\begin{equation}
[1+(1-q)kx]\frac{dN}{dx}=-\lambda N,
\end{equation}
that can also be considered as an equation that describes a non constant
death probability, that is, a death probability that depends on the
dose: 
\begin{equation}
p(x)=\frac{\lambda}{[1+(1-q)kx]}.\label{eq:cell death prob 1}
\end{equation}

The solution is 
\begin{equation}
\frac{N(x)}{N_{0}}=[1+(1-q)kx]^{-\frac{\lambda}{k(1-q)}}=[e_{q}(kx)]^{-\lambda/k}.
\end{equation}

Now, if we redefine variables as $k(1-q)=-1,$ $-\frac{\lambda}{k(1-q)}\equiv\gamma,$
that is nothing more than the model for survival factor in Ref. \citep{Sotolongo-PRL},
$F_{S}=(1-x)^{\gamma}.$ This model has excellent experimental proof
in the literature \citep{Sotolongo-PRL,Brouers-Sotolongo}. The linear-quadratic
(LQ) model is recovered in the limit $q\rightarrow1$ up to second
order in a Taylor series expansion.

\section{Survival Probability}

In this section, we make some probabilistic arguments in order to
reinforce the consistence of the models\citep{Sotolongo-PRL,Brouers-Sotolongo}.

Consider the cell death probability due to radiation absorbed as

\begin{equation}
p(x)=\frac{\lambda}{(1-x)},\label{eq:Probability}
\end{equation}
that is the probability given in eq.\ref{eq:cell death prob 1}, with
$k(1-q)=-1$.

Consider now the Escort Probability, in such a way that it is defined
in NESM as

\begin{equation}
P(x)=\frac{p^{r}(x)}{\int p^{r}(x)dx},
\end{equation}
with normalized probability condition 
\begin{equation}
\intop_{0}^{1}P(x)dx=1.
\end{equation}
Where the entropic index is written here as $r$, in order to distinguish
from the index $q$ used in the previous section.

A simple calculation shows that, using eq. \ref{eq:Probability},
the escort probability $P(x)$ can be written as

\begin{equation}
P(x)=(1-r)(1-x)^{-r}.
\end{equation}
If we call $1-r=\gamma,$ this lead to the a death probability that
depends on the dose as $\gamma(1-x)^{\gamma-1}$and is similar to
the result of Ref. \citep{Brouers-Sotolongo}, but here for radiation
only. The main problem in this attempt to connect escort probability
with the approach of structural derivatives is that the $r$ parameter
here must be negative to give $\gamma>1.$

Independent of this, lets try to obtain the factor $F_{S}$ in this
context of NESM.

The survival factor can be seen as the probability difference from
the probability of whole conformational space and the probability
of cell death by radiation, that is,

\begin{eqnarray}
F_{S} & = & \left[1-\intop_{0}^{x}P(x)dx\right].
\end{eqnarray}
This can be rewritten as 
\begin{equation}
F_{S}=\intop_{x}^{1}P(x)dx.
\end{equation}
Again, a simple calculation leads to

\begin{equation}
F_{S}=(1-x)^{1-r}=(1-x)^{\gamma},
\end{equation}
that is our the previous result for the survival factor.

Note that, if we redefine the $r$ parameter as $1-r=\frac{2-q}{1-q}=\gamma>1,$
the result seems to be now compatible with that of Refs. \citep{Sotolongo-PRL,Brouers-Sotolongo}.

\section{Conclusions and Outlook}

In this contribution, we introduced, for the first time, the mathematical
tool of structural derivatives to model the survival factor for tissues
irradiation.

The models are analogous to the nuclear decay problem, that takes
the time, $t$, as the independent variable; here, instead, our independent
variable is the applied dose, $x.$

We claim that the deformations of the derivative are the key to model
complex systems; particularly here, we emphasize and focus on life,
medical and biological science. Here, we adopt the local forms of
derivative, called structural derivatives. However, we think that
nonlocal forms of derivative, like fractional calculus, can help modeling
too, but with much more computational and algebraic cost. We think
that our model could help researchers to better understand models
and dynamics, without excessive heuristics in the formulation.

As outlooks, we shall develop models with other forms of structural
derivatives, in an attempt to achieve survive factors, that could
be tested for the adequate kind of tumor cells. Also, the structural
derivative may be applied to model earthquakes \citep{Sotolongo- Earthquakes1}. 
\begin{acknowledgments}
The authors would like to thank J. A. Helayël-Neto for a critical
reading of this article.

This work was performed under support of PRODEP project DSA/103.5/15/986
from SEP, Mexico.\end{acknowledgments}

\end{document}